\documentclass[prl,aps,twocolumn,showpacs,preprintnumbers,amsmath,amssymb]{revtex4}
\usepackage{bm}
\usepackage{epsfig}

\begin{document}

\preprint{}

\title{Orbital-dependent metamagnetic response in Sr$_4$Ru$_3$O$_{10}$}
\author{Y. J. Jo,$^1$ L. Balicas,$^1$ N. Kikugawa,$^{2,\ast}$ E. S. Choi,$^1$ K. Storr,$^3$ M. Zhou,$^4$ and Z. Q. Mao$^4$ }

\affiliation{$^1$National High Magnetic Field Laboratory, Florida
State University, Tallahassee-FL 32306, USA}
\affiliation{$^2$School of Physics and Astronomy, University of St. Andrews, St. Andrews, Fife KY16 9SS, UK}
\altaffiliation{Present address: National Institute for Materials Science, Sengen, Tsukuba 305-0047, Japan}
\affiliation{$^3$Department of Physics, Prairie View A\&M University, Texas 77446-0519, USA}
\affiliation{$^4$Department of Physics, Tulane University, New Orleans, Louisiana 70118, USA}

\date{\today}%
\begin{abstract}
We show that the metamagnetic transition in Sr$_4$Ru$_3$O$_{10}$ bifurcates into two transitions as the field is rotated away from the conducting planes.
This two-step process comprises partial or total alignment of moments in ferromagnetic bands followed by an itinerant metamagnetic transition whose critical field increases with rotation.
Evidence for itinerant metamagnetism is provided by the Shubnikov-de Hass effect which shows a non-trivial evolution of the geometry of the Fermi surface
and an enhancement of the quasiparticles effective-mass across the transition. The metamagnetic response of Sr$_4$Ru$_3$O$_{10}$ is orbital-dependent and involves ferromagnetic and metamagnetic bands.

\end{abstract}

 \maketitle

\section{Introduction}

Metamagnetism is usually understood as a rapid increase of the magnetization of a given system in a narrow range of magnetic fields.
In a spin localized picture, it would correspond to the field-induced suppression of, for instance, antiferromagnetic order via a
spin-flop or the subsequent spin-flip transition \cite{review}. In itinerant systems, metamagnetism is explained either in terms of the field-induced spin-polarization
of the Fermi surface and concomitant field-tuned proximity of the Fermi level to a van-Hove singularity \cite{binz} or in terms of the suppression of antiferromagnetic correlations \cite{review}.
So far, both itinerant metamagnetic scenarios have been discussed within a single band picture \cite{binz,review}.

The ruthenates, Sr$_3$Ru$_2$O$_7$ in particular, were reported to display complex metamagnetic behavior, including the possibility of quantum-criticality by
tuning to zero temperature the end point of a first-order metamagnetic transition \cite{robin}. At low temperatures and in high purity samples, this first-order line
bifurcates into two first-order transitions that define the boundary of a new phase \cite{santiago} emerging at the quantum-critical end point
which is claimed to result from the coupling of lattice fluctuations to a Fermi surface
instability \cite{green}. Complex and highly anisotropic metamagnetic behavior has also been recently reported in
the tri-layered  Sr$_{4}$Ru$_{3}$O$_{10}$ compound \cite{cao} which is a structurally distorted ferromagnet: the RuO$_6$ octahedra in the outer two layers of each triple
layer are rotated by an average of $5.6^{\circ}$ around the c-axis, while the octahedra of the inner layers are rotated in the opposite sense by an average of $11.0^{\circ}$ \cite{crawford}. It displays
a Curie temperature $T_{c} \sim 100 $ K (see upper panel of Fig. 1 which shows the temperature dependence of both the heat capacity $C$ normalized respect to the temperature $T$ as well as the low field magnetization $m$)
and a saturated moment $ S \gtrsim 1 \mu_B$ per Ru$^{4+}$ ion, which Raman spectroscopy suggests to be \emph{localized} in the Ru site \cite{gupta} and
directed essentially along the c-axis \cite{crawford,cao}. The ferromagnetic (FM) transition is followed by an additional broad peak in the magnetic susceptibility at $T_M \simeq 50 $ K in flux grown samples \cite{crawford,cao}
($T_M \simeq 70$ K in floating zone grown crystals) which is claimed to result from  the ``locking" of the moments into a
canted antiferromagnetic configuration \cite{gupta}. Nevertheless, neutron scattering experiments indeed find the moments lying in the ab-plane but coupled ferromagnetically \cite{bao}.
Furthermore, the behavior of the magnetic susceptibility $m/H$, shown in Fig. 1, displaying a ferromagnetic transition followed by a broad maximum as the temperature is lowered, is
consistent with the reentrance of the paramagnetic state at lower temperatures, a conventional behavior of itinerant metamagnetic systems close to a van Hove singularity \cite{review,binz}.
Here we show that Sr$_{4}$Ru$_{3}$O$_{10}$ displays clear evidence for metamagnetic behavior involving simultaneously several of the bands associated with the $t_{2g}$ orbitals.

\section{Results and Discussion}

Figure 2 (a) displays the magnetization $m$ of a Sr$_{4}$Ru$_{3}$O$_{10}$ single-crystal measured with a commercial SQUID magnetometer as a function of the field $H$ at $T=5$ K and for several values of
the angle $\theta$ between $H$ and the c-axis. When $H$ is applied along an in-plane direction, ($\theta = 90^{\circ}$) it exhibits a sharp and hysteretic metamagnetic transition at a critical field $H_{MM} \sim 2.5$ T \cite{crawford,cao}.
In high quality single crystals the observed hysteresis exhibits ultra-sharp steps in the resistivity
resulting from domain formation \cite{mao}. But as $\theta$ decreases, as indicated by the vertical arrows, the transition clearly occurs via an anisotropic two-step process.
A first step in $m$ is observed at $\sim 1$ T and moves to lower fields as $\theta$ decreases leading at high angles and above a field of just $0.2$ T, to a saturation value $m \simeq 1.3$ $\mu_B$ per Ru.
Notice that according to Hund's rule, the placement of the four Ru$^{+4}$ carriers on the three $4d$ $t_{2g}$ orbitals leads to an effective moment $S = 1$ or $2$ $\mu_B$ per Ru. If
the lattice distortions in Sr$_{4}$Ru$_{3}$O$_{10}$ do not lift the degeneracy among the three $t_{2g}$ orbitals, one would expect each orbital to contribute 2/3 of a $\mu_B$ per Ru to the total moment.
Thus a saturation value of $2 \times 2/3 \simeq 1.3$ $\mu_B$ per Ru, suggests the polarization of the moments in \emph{two} of the $t_{2g}$ orbital networks.
While a second step is observed in the magnetization when $\theta < 90^{\circ}$ and moves to much higher fields as $\theta$ decreases producing considerable hysteresis.
This behavior is confirmed by measurements of the magnetic torque $\overrightarrow{\tau} = n \overrightarrow{m} \times \overrightarrow{H}$ (where $n$ is the total number of Ru atoms) measured by using a thin film CuBe cantilever
(5/1000 of an inch in thickness) at $T = 4.2$ K shown in  Fig. 2 (b). Notice how sharp and hysteretic this second metamagnetic transition is. Its sharpness is a very strong indication of the high quality of our samples.
Remarkably, at low angles this second metamagnetic transition occurs under a sizeable component of $H$
applied along the inter-plane direction, or with the moments basically aligned along the c-axis by the first transition.

\begin{figure}
\begin{center}
\epsfig{file=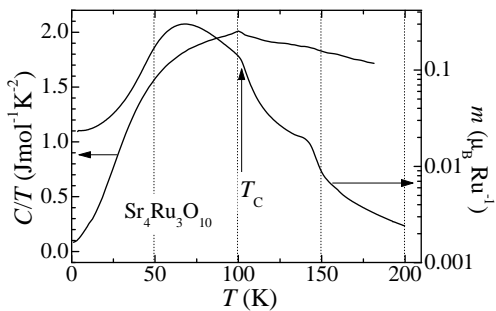, width=6.5 cm} \caption{Heat capacity $C$ normalized respect to temperature $T$ and magnetization $m$ (for a field of 500 Gauss applied
along an in-plane direction) for a Sr$_4$Ru$_3$O$_{10}$ single crystal and as a function of temperature $T$. Only one single anomaly, the Curie temperature, is observed in \emph{both} traces at $T_c \sim 100$ K.
No indications of sub-phases is found.}
\end{center}
\end{figure}

In order to understand this complex metamagnetic behavior, particularly the high field metamagnetic transition, we performed electrical transport measurements
at very high magnetic fields and low temperatures in two batches of single crystals, one having a residual resistivity $\rho_0 \simeq 6 \mu \Omega$cm, where we performed
inter-plane transport measurements, and a second one displaying a $\rho_0$ between 1.5 and 2 $\mu \Omega$cm used mainly for in-plane transport studies.
Our goal is two-fold: i) to measure the evolution of the geometry of the Fermi surface and of the quasiparticles effective mass across the high field metamagnetic transition via the Shubnikov-de-Haas effect,
to explore the possibility of itinerant metamagnetism, ii) to determine the geometry of the Fermi surface in order to stimulate band structure calculations which could clarify the
possible existence of localized bands.

\begin{figure}
\begin{center}
\epsfig{file=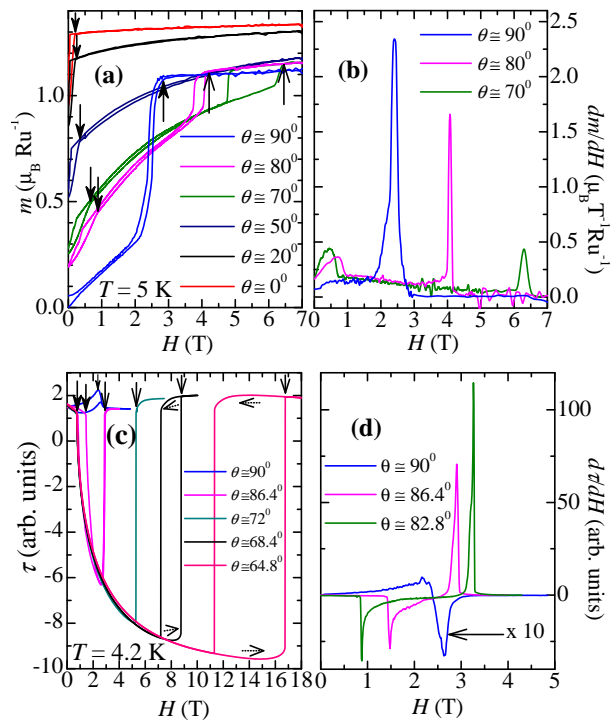, width=8 cm} \caption{(a)Magnetization $m$ as a function of field $H$ for several angles $\theta$ between $H$ and the inter-plane c-axis. Arrows indicate
the metamagnetic transitions. (b) The derivative of $m$ respect to the field $\partial m/ \partial H$ and as a function of field $H$ displaying peaks at the metamagnetic transitions.
(c) Magnetic torque $\tau$ as a function of $H$ at 4.2 K. Vertical arrows indicate the metamagnetic transitions, dotted arrows indicate increasing or decreasing field sweeps.
(d) The derivative of the torque $\tau$ respect to the field $\partial \tau / \partial H$ and as a function of field $H$.}
\end{center}
\end{figure}
\begin{figure}[htb]
\begin{center}
\epsfig{file=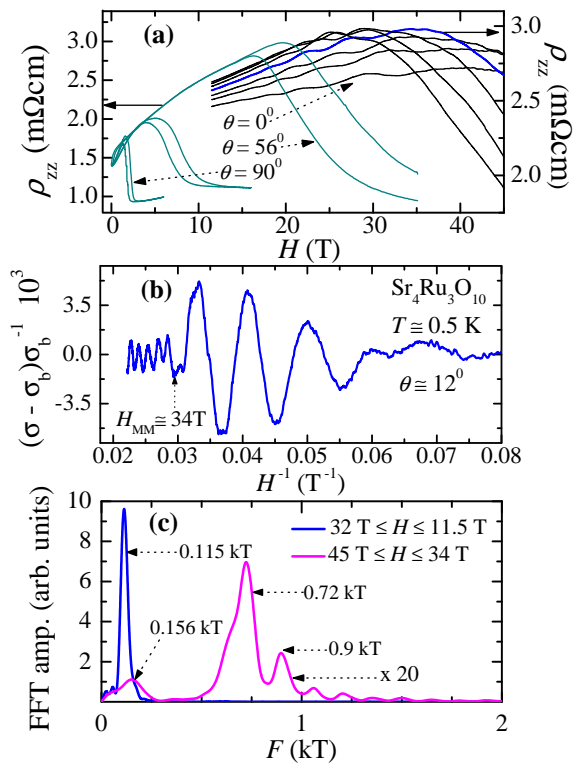, width = 7.5 cm} \caption{(a) The
inter-plane resistivity $\rho_{zz}$ for two Sr$_4$Ru$_3$O$_{10}$ single
crystals as a function of magnetic field $H$ at $T \simeq 0.5$ K
and for several angles $\theta$ between $H$ and the inter-plane
c-axis. Notice both the presence of Shubnikov de Haas oscillations and
the rapid decrease in $\rho_{zz}$ when the metamagnetic transition is crossed
at higher angles. (b) The SdH signal $(\sigma - \sigma_{b}) \sigma_{b}^{-1}$ where $\sigma_{b} = 1/ \rho_{b}$ is the background conductivity as a function $H^{-1}$
for the blue trace at $\theta = 12 \pm 2 ^{\circ}$ in (a). (c) The fast Fourier transform (FFT) of the SdH signal in (b) for two ranges in field, i.e, above and below the metamagnetic transition.}
\end{center}
\end{figure}

Our crystals were grown by a floating-zone (FZ) technique \cite{mao_crystals}. Crystals selected for the measurements were characterized by x-ray diffraction, heat capacity, and magnetization measurements and were found to
be pure Sr$_{4}$Ru$_{3}$O$_{10}$. For instance, in Fig. 1, only one weak anomaly is seen in $C/T$ at $T_c \simeq 100$ K indicating the onset of ferromagnetism. No clear anomaly is seen that one could associate with the proposed spin
canting transition \cite{gupta} or with the existence of inclusions of other phases. The absence of inclusions, as seen for example in self-flux grown samples \cite{yan}, is confirmed by measurements of the magnetization $m$
as a function of $T$ for a field of $H = 500$ Gauss applied along an in-plane direction, also shown in Fig. 1. One sees only a couple of anomalies, one at $T_c$ and a second one at 140 K which does not leave any clear signature in $C/T$.
Inclusions of SrRuO$_3$ would produce an anomaly at $T_c = 160$ K and would lead to a rapid increase in $m$ when low fields are applied along the ab-plane, contrary to what is seen respectively in Figs. 1 and 2(a).
Electrical transport measurements were performed by using conventional four-probe techniques in conjunction with a single axis-rotator inserted in a $^3$He cryostat. High magnetic fields up to 45 T were provided
by the NHMFL.

Figure 3 (a) shows the inter-plane resistivity $\rho_{zz}$ for two Sr$_{4}$Ru$_{3}$O$_{10}$ single crystals (typical dimensions $0.5 \times 0.5 \times 0.3$ mm$^{3}$) at $T \simeq 0.6$ K as a function of $H$ and for
several angles $\theta$ between $H$ and the inter-plane c-axis. Notice both, the oscillations in $\rho_{zz}$ or the Shubnikov de Haas (SdH) effect, and the marked negative magnetoresistivity \emph{emerging
from the second metamagnetic transition} seen at higher fields. For $\theta \simeq (12 \pm 2)^{\circ}$, for example, the metamagnetic transition field is $\sim$ 34T (blue trace),
and is displaced to fields beyond 45T as the field is aligned along the c-axis.
Figure 3 (b) displays the oscillatory component or the SdH signal (defined as $(\sigma - \sigma_{b})/ \sigma_{b}$
where $\sigma = 1/\rho$ and $\sigma_b$ is the inverse of the background resistivity) of the trace shown in (a) ($\theta \simeq (12 \pm 2)^{\circ}$) as a function of the inverse field $H^{-1}$.
Both the amplitude and the frequency of the oscillatory component changes drastically at $H_{MM} \sim 34$ T as quantified by the Fast Fourier Transform of the SdH signal taken in
two limited field ranges, i.e., from 11.5 to 32 T (or below $H_{MM}$) in blue and from 34 to 45 T (or above $H_{MM}$) in magenta, both shown in Fig. 3(c). The peak seen at $F = 115$ T \cite{cao} is completely
suppressed by the transition implying the reconstruction of the Fermi surface.
\begin{figure}[htb]
\begin{center}
\epsfig{file=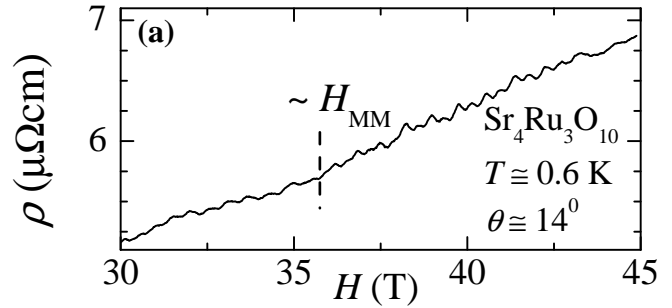, width = 8.6 cm}
\caption{In-plane resistivity $\rho(H)$ as a function of field $H$ in a limited field range for $T= 0.6$ K and $\theta = 14^{\circ}$ }
\end{center}
\end{figure}

\begin{figure}[htb]
\begin{center}
\epsfig{file=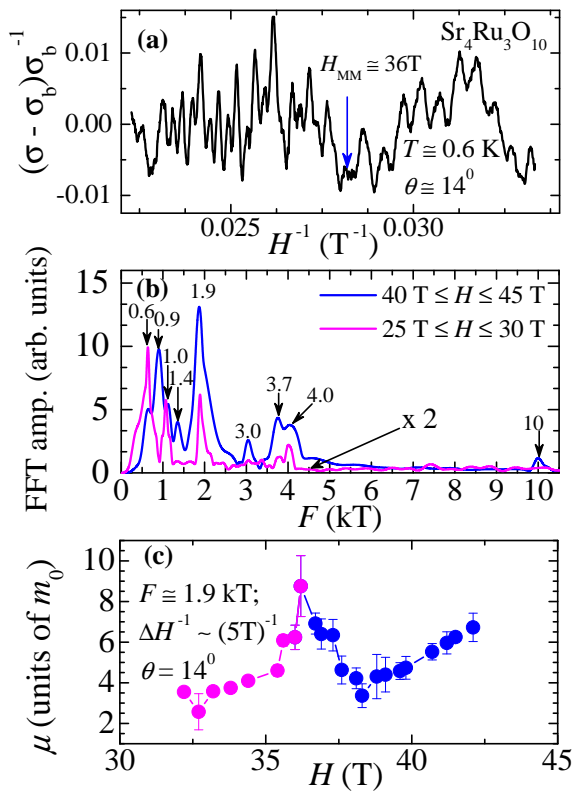, width = 7.5 cm} \caption{(a) Oscillatory component of the in-plane resistivity $\rho$ for a Sr$_4$Ru$_3$O$_{10}$ single
crystal as a function of $H$ at $T \simeq 0.6$ K
and for $\theta = (14 \pm 2)^{\circ}$ between $H$ and the inter-plane
c-axis. (b) The FFT spectrum of the SdH signal shown for two limited field ranges, from 25 to 30 T (in magenta)
and from 40 to 45 T (in blue). Arrows indicate the frequency of the main peaks in kT.
(c) The magnetic field dependence of the effective mass associated with the FFT peak at $F = 1.9$ kT.
The temperature dependence of the FFT spectra was taken within a window in $\Delta(H)^{-1} \simeq (5T)^{-1} $.}
\end{center}
\end{figure}
A more thorough investigation of the effect of the transition on the Fermi surface is given by measurements of the in-plane magnetoresistivity $\rho(H)$  in
the highest quality single crystals of Sr$_4$Ru$_3$O$_{10}$ (dimensions $0.6 \times 0.6 \times 0.04$ mm$^{3}$) currently available.
As an example of typical raw data we show in Fig. 4 the in-plane resistivity as a function of the external field in a limited range, i.e., between 30 and 45 T at
$T = 0.6$ K, and for an angle $\theta \simeq 14^{\circ}$. The SdH oscillations are clearly visible although the metamagnetic transition is barely perceptible.
Notice how Sr$_4$Ru$_3$O$_{10}$ is extremely anisotropic, $\rho_{zz} / \rho \sim 1000$,  consequently one should expect to observe nearly two-dimensional Fermi surface sheets.
Figure 5 (a) shows the SdH signal obtained from the trace of $\rho(H)$ in Fig. 4. According to Fig. 2 (a), at this angle under a field of 7 T the net moment is nearly saturated to a value of $\sim 1.3$ $\mu_{B}$. While
from our previous angular study in Fig. 3 (a), at this angle one expects the metamagnetic transition to happen in the neighborhood of 35 T.
One can clearly see that the oscillatory pattern changes in the neighborhood of 36 T.
In Fig. 5 (b) we display the FFT spectrum of the SdH signal for two ranges of magnetic field, i.e., from 25 to 30 T or clearly below the metamagnetic transition (in magenta), and from 40 to 45 T where the
geometry of the Fermi surface is closer to being stable (in blue). Larger frequencies such as the one at $\sim 2$ kT and those around 4 kT, remain unaffected
by the transition. However the entire spectral weight below 2 kT is shifted towards higher frequencies.
This situation is somewhat similar to that reported in the bi-layered compound Sr$_3$Ru$_2$O$_7$ where the FFT spectrum reveals additional structure and small shifts in the frequency
of the main peaks across its metamagnetic transition which are claimed to result from the spin-splitting of the Fermi surface \cite{borzi}.
In Figure 5 (c) we plot the effective mass $\mu$ in units of free electron mass $m_0$, associated with the peak observed at $F = 1.9$ kT and as a function of $H$.
Here we measured $\rho$ at different temperatures and extracted the corresponding FFT spectrum within a narrow window in $H^{-1} \sim (5T)^{-1}$.
The evolution of the amplitude of a given peak in frequency as a function of temperature was fitted to the usual Lifshitz-Kosevich expression $x/sinhx$ to
extract $\mu$. Notice how the value of $\mu$ spikes at $\sim 36$ T, a behavior similar to that of  Sr$_3$Ru$_2$O$_7$ and which is claimed to result
from the fluctuations emerging from a metamagnetic quantum-critical end point \cite{borzi}. A metamagnetic quantum-critical end point in the context
of itinerant metamagnetism has also been predicted for Sr$_{4}$Ru$_{3}$O$_{10}$ \cite{binz} but more work is needed to clarify its existence.
In any case, the enhancement of effective mass is also consistent with the scenario of a Zeeman-split
Fermi surface that crosses a nearby van-Hove singularity as the field increases, i.e., conventional itinerant metamagnetic scenario.

In Sr$_4$Ru$_3$O$_{10}$ the amplitude of the oscillations is quickly damped as $\theta$ increases. However, we were able to fit (not shown here) the angular dependence of the main peaks
corresponding respectively to 1.9, 3.95, and 9.96 kT, and which are associated with largest cross-sectional areas of the Fermi surface, to the expression $F = F_0 / \cos \theta$, indicating their two-dimensional character.
Instead, lower frequencies such as the $\sim 1$ kT display a linear dependence. Although one could naively expect Sr$_4$Ru$_3$O$_{10}$ to display a nearly three-dimensional Fermi surface due to its structural proximity to
the infinite layered compound SrRuO$_3$, the two-dimensional character of its main FS sheets agrees well with its highly anisotropic electrical transport properties.
Finally, the effective masses associated with these orbits averaged over our entire field range are $4.4 \pm 0.5$, $5.6 \pm 0.3$, $6 \pm 1$ and $21 \pm 6$ for the 1.02, 1.9. 3.95 and 9.96 kT frequencies, respectively.

\section{Conclusions}
In summary, we have shown that the metamagnetic behavior observed in Sr$_4$Ru$_3$O$_{10}$ \emph{bifurcates into two transitions} as the external field rotates away from the conducting planes.
A first step observed in the magnetization, which results from the alignment of moments in ferromagnetic bands,
moves to fields well below 1 T and leads to a large polarized moment of 1.3 $\mu_B$ per Ru when an external field is applied nearly along the c-axis. While a second magnetization step
shows a pronounced angular dependence and moves to fields beyond 45 T when it is applied nearly along the inter-plane axis.
The enhancement of the quasiparticles effective mass and the change in the geometry of the Fermi surface at the transition, are indications of its itinerant metamagnetic character.
We conclude that metamagnetism in Sr$_4$Ru$_3$O$_{10}$ is an orbital-dependent process involving both the alignment of moments/domains in ferromagnetic bands that are either itinerant or localized, and
the polarization of an itinerant band that is in close proximity to a van-Hove singularity. One possible scenario is that the external field applied along the ab-plane aligns the
FM moments/domains producing a large internal field that couples to the lattice via magneto-elastic coupling. The concomitant changes in lattice constants could lead to the itinerant metamagnetic response.
But as the field is rotated away from an in-plane direction, the metamagnetic transition would occur whenever the sum of the in-plane components of the internal field (which progressively tilts towards the c-axis) and the external
field is equal to the sum of both fields when $H\|$ ab-plane. In any case, the complex and simultaneous involvement of several bands in the magnetic response
has yet to be considered by theoretical treatments of metamagnetism \cite{review, binz}. We owe this complex physical behavior to the multi-band nature of this compound.
Orbital-dependent superconductivity in Sr$_2$RuO$_4$ \cite{daniel} and the proposed orbital-selective Mott transition in the
Ca$_{2-x}$Sr$_{x}$RuO$_4$ system \cite{anisimov} are other examples of possible multi-band behavior in the ruthenates.
Band structure calculations for Sr$_4$Ru$_3$O$_{10}$ would be highly desirable since a comparison with our experimental Fermi surface
determination could clarify the proposed existence of localized bands \cite{gupta}.

\section{Acknowledgements}
We acknowledge useful discussions with P. B. Littlewood and V. Dobrosavljevic. The NHMFL is supported by NSF through NSF-DMR-0084173 and the State of Florida.  YJJ acknowledges support from the NHMFL-Schuller
program. Z.Q. Mao is a ``Cottrell Scholar" and acknowledges the Louisiana Board of Regents fund through LEQSF(2003-06)-RD-A-26 and the pilot fund NSF/LEQSF(2005)-pfund-23
as well as support from the NHMFL's visiting scientist program (VSP). KS also acknowledges the NHMFL VSP. LB was supported by the NHMFL in-house program.

\end{document}